\documentclass[11pt]{article} 
\usepackage{amsmath,amsfonts,graphicx,mathrsfs,tikz,rotating} 
\usetikzlibrary{shapes.geometric,shapes.arrows,decorations.pathmorphing}
\usetikzlibrary{matrix,chains,scopes,positioning,arrows,fit}

\newcommand{\be}{\begin{align}}
\newcommand{\ee}{\end{align}}
\newcommand{\bear}{\begin{eqnarray}}
\newcommand{\eear}{\end{eqnarray}}

\newcommand{\ba}{\begin{array}}
\newcommand{\ea}{\end{array}}

\newcommand{\diag}{\textrm{diag}}

\begin{document}
\title{Eigenvalue spectra of asymmetric random matrices for multi-component neural networks}
\author{ Yi Wei \\
Cold Spring Harbor Laboratory, Cold Spring Harbor, New York 11724, USA}
\date{}
\maketitle
\vspace*{-10mm}
\abstract{
This paper focuses on large neural networks whose synaptic connectivity matrices are randomly chosen 
from certain random matrix ensembles. The dynamics of these networks can be characterized by the eigenvalue 
spectra of their connectivity matrices. In reality, neurons in a network do not necessarily behave 
in a similar way, but may belong to several different categories. The first study of the spectra 
of two-component neural networks was carried out by Rajan and Abbott. In their model, neurons are either 
'excitatory' or 'inhibitory', and strengths of synapses from different types of neurons have Gaussian 
distributions with different means and variances. A surprising finding by Rajan and Abbott is that the 
eigenvalue spectra of these types of random synaptic matrices do not depend on 
the mean values of their elements. In this paper we prove that this is true even for a much more 
general type of random neural network, where there is a finite number of types of neurons, and their 
synaptic strengths have correlated distributions. Furthermore, using the diagrammatic techniques, we 
calculate the explicit formula for the spectra of synaptic matrices 
of  multi-component neural networks.

\section{Introduction}
In neuroscience, interconnections of neurons are often represented by synaptic matrices whose 
elements are drawn from a certain random matrix ensemble \cite{ra,scs}. Knowing the distribution of 
eigenvalues of these random matrices is very important in studying spontaneous activities and 
evoked responses of the network. To calculate the eigenvalue distribution of these matrices, 
it is often necessary to work with asymmetric (non-hermitean) random matrix theory, which has been 
successfully applied to many fields of physics and interdisciplinary sciences, e.g. the phase 
diagram of QCD \cite{ste,tilo}, nuclear decay and resonances in multichannel chaotic scattering 
\cite{jac} and neural networks\cite{ra,scs,som}. 

A prominent result of asymmetric random matrix theory is Girko's circle law \cite{girko}. In its 
variation with partial symmetry, the circle becomes an ellipse \cite{som}. These classic results, 
however, can not be directly applied to realistic neural network models where neurons do not behave 
in the same way \cite{ra,she,hol}. Assume there are $N$ number of neurons and let $W$ be the synaptic 
matrix. In the model of Rajan and Abbott \cite{ra}, there are $fN$ number of neurons which are 'excitatory', 
and all others are 
'inhibitory'. To model this neural network, elements in $fN$ columns of $W$ are sampled from  a 
Gaussian distribution with mean $\mu_E/\sqrt{N}$ and variance $\sigma^2_E/N$ and elements in the remaining 
$(1-f)N$ columns of $W$ are Gaussian variables with mean $\mu_I/\sqrt{N}$ and variance $\sigma^2_I/N$. 
Therefore, the synaptic matrix has the structure $W=J\Lambda+M$, where $J$ is drawn from the real Ginibre 
ensemble \cite{gin} such that $\langle J_{ij}\rangle\!=\!0$, $\langle J^2_{ij}\rangle\!=\!1/N$, and 
$\Lambda=\diag(\sigma_E I_{fN},\sigma_I I_{(1-f)N})$, where $I_{fN}$ and $I_{(1-f)N}$ are identity matrices 
of dimension $fN$ and $(1-f)N$, respectively. $M$ is a constant matrix whose elements are the mean 
strength of the synapses. Since there are two types of synapses, every row of $M$ is identical and in each row, 
the first $fN$ elements are equal to $\mu_E/\sqrt{N}$ and the remaining $(1-f)N$ elements equal to 
$\mu_I/\sqrt{N}$. In particular, $M$ is chosen to be in a 'balanced' situation 
such that $f\mu_E+(1-f)\mu_I=0$ \cite{newn,miller}. To confine eigenvalues inside a unit circle, a second 
constraint \cite{ra} is introduced which requires that the strengths of the synapses attached to each neuron 
independently sum to zero. It is found in \cite{ra} that, in the limit $N\to\infty$, modifying the mean 
strengths of excitatory and inhibitory synapses has no effect on the eigenvalue spectra of the synaptic 
matrices. Therefore, the spectrum of $W$ is identical to that of $J\Lambda$. 

It is natural to wonder why the 'mean' strength matrix $M$ has no effect on the spectra. Moreover, in real 
biological neural systems, several different types of neurons may connect each other to form a multi-component 
network \cite{she,hol}. Distributions of synaptic strengths of different types of neurons 
are distinct \cite{bb} and could be non-Gaussian \cite{bb,mitya}. Dynamics of this network therefore depend 
on properties of each type of neuron. It is interesting to find whether or not the eigenvalue density of this 
type of networks depends on the mean value of each individual type of synapse. These questions are addressed in 
section 2. One of the main results of this paper shows that even without the second constraint in \cite{ra} 
and when synaptic strengths have certain non-Gaussian distributions, the spectrum of the network still 
does not depend on the mean synaptic strengths.

Aside from biological motivations, the eigenvalue problem of random plus fixed matrices has been a 
research topic in both random matrix theory and condensed matter physics for a long time \cite{bz,zeeadd,tt,ss}. 
A different point of view of the problem in this paper is: 
how is the density function of large random matrix $J\Lambda$ perturbed by the rank-1 constant matrix $M$. 
Note that random matrix $J\Lambda$ is not of Wigner type, nor are its elements independently and identically 
distributed (iid). Therefore this paper provides new results to similar problems studied in \cite{tt,ss}.

Furthermore, finding the eigenvalue density of random matrices of the form $J\Lambda$, where $J$ is drawn from 
a random matrix ensemble and $\Lambda$ is a fixed matrix, has been an interesting topic in random matrix theory 
and mesoscopic physics \cite{kz}. When $J$ is drawn from the circular unitary ensemble (CUE), an exact result 
is given in \cite{wei}, and the large-N limit is calculated in \cite{bog}. In section 3, we calculate the density 
function of $J\Lambda$ where $J$ belongs to the real Ginibre 
ensemble using the method introduced in \cite{man,zee}. 
Discussion and remarks are made in the last section.

\section{Synaptic strength of non-Gaussian distributions}


Let the $N\times N$ dimensional real matrix $W$ be the synaptic matrix of an $N$-neuron network. Assume there 
are $m$ types of neurons and the i-th type of neuron has a population of $f_iN$, $\sum_{i=1}^mf_i=1$. Define 
a constant diagonal matrix 
\begin{align}\label{eq:la}
\Lambda=\mathrm{diag}(\sigma_1I_{f_1N},\dots,
\sigma_mI_{f_mN})>0,
\end{align} 
where $I_{f_iN}$ is the $f_iN$-dimensional identity matrix. Let $v$ be an $N$-dimensional row vector with the 
following form,
\begin{align}\label{eq:v}
v=({\underbrace{\mu_1,\dots, \mu_1}_{f_1N}}, \underbrace{\mu_2,\dots,\mu_2}_{f_2N},\dots,
\underbrace{\mu_m,\dots,\mu_m}_{f_mN} ),
\end{align}
where $\mu_i$ is the mean strength of the synapses from neurons of the i-th type. Define the $N\times N$ 
dimensional 'mean' matrix $M$, whose rows are all equal to $v$. The synaptic matrix $W$ in our model takes the form 
\begin{align}\label{eq:w}
W=J\Lambda+M,
\end{align} 
where $J$ is an $N\times N$ dimensional real random matrix drawn from the ensemble 
\begin{align}\label{eq:ensemble}
P(J)=\frac{1}{Z}\exp(-N\mathrm{tr} V(JJ^T)),
\end{align}
where $V$ is an arbitrary function and $Z$ is the normalization constant. The case $V(x)=x/2$ corresponds to 
Ginibre ensemble where elements of $J$ are statistically independent Gaussian variables. By symmetry, the mean 
of $J_{ij}$ vanishes, 
$\langle J_{ij}\rangle=0$.  The variance of $J_{ij}$ is determined by $V$, which is normalized to 
be $1/N$, i.e. $\langle J^2_{ij}\rangle=1/N$.  

Synaptic matrix $W$ defined in Eq.\eqref{eq:w} has $m$ column blocks. Each block corresponds to one type of 
neuron. By construction, elements of $W$ have the following statistical properties, 
\begin{align}\label{eq:wstat}
\mathrm{Var}(W_{ij})=\Lambda_{jj}^2/N \ \ \mathrm{and}\ \ \langle W_{ij}\rangle=v_j,
\end{align}
i.e. the i-th type of synaptic strength has variance $\sigma_i^2/N$ and mean $\mu_i$. The matrix $M$ has a 
similar column block structure as the one defined in \cite{ra}. Following \cite{ra}, we also choose to put the 
synapses at the 'balanced' situation, i.e.
\begin{align}\label{eq:bal}
\sum_{i=1}^{m}f_i\mu_i=0.
\end{align}

We want to know the eigenvalue density $\rho(x,y)$ of $W$ in the limit $N\to\infty$, with fixed $f$'s. Let 
$z=x+\mathrm{i}y$, the density $\rho(x,y)$ is related to the Green's function $G_W(z,\bar{z})$ as 
\begin{align}\label{eq:rho}
\rho(x,y)=\frac{1}{\pi}\frac{\partial}{\partial\bar{z}}G_W(z,\bar{z}),\ \ {\mathrm{where}}\ \  
G_W(z,\bar{z})=\frac{1}{N}\left\langle\mathrm{tr}_N\frac{1}{\ z-W \ }\right\rangle_J.
\end{align} 
In the above formula, $\langle\cdots\rangle_J$ means averaging over the ensemble Eq.\eqref{eq:ensemble} of 
matrix $J$. We write the Green's function as $G_W(z,\bar{z})$ to emphasize it is not analytic on a 
2-dimensional region of the $(x,y)$-plane, more details can be found in \cite{zee}. This region is called 
the support of the density function since on which we have $\frac{\partial}{\partial \bar{z}}G_W(z,\bar{z})\ne0$.
Since we will be dealing with both $N\times N$ and $2N\times2N$ dimensional matrices, to remove ambiguity, 
we use $\mathrm{tr}_N$ as the trace operator for $N\times N$ matrices. 
We will work on asymmetric random matrices with the methods introduced in \cite{man,zee}. For consistency, 
we adopt the notation convention of \cite{man} in the remaining of this paper. Define a 
$2N\times2N$ dimensional matrix 
\begin{align}\label{eq:z}
\mathcal{Z}=
\left( \begin{array}{cc} z & \lambda \\ \lambda & \bar{z}  \end{array}\right).
\end{align} 
For asymmetric matrix $W$, define the resolvent (matrix valued Green's function) $\mathcal{G}_W$ 
\cite{man,zee} as  
\begin{align}\label{eq:G}
\mathcal{G}_W(\mathcal{Z})=\left(\begin{array}{cc} \mathcal{G}_1 & \mathcal{G}_2 \\
\mathcal{G}_3 & \mathcal{G}_4\end{array}\right)=
\left\langle
\left[ \mathcal{Z}- \left(\begin{array}{cc} W &  \\  & W^T  \end{array}\right)\right]^{-1}_{2N\times2N}
\right\rangle_J.
\end{align} 
Introducing the self-energy $\Sigma_W$, we have
\begin{align}\label{eq:self}
\mathcal{G}_W(\mathcal{Z})
=\frac{1}{\mathcal{Z}-\Sigma_W}.
\end{align} 
The Green's function in Eq.\eqref{eq:rho} can be found from $\mathcal{G}_W$ \cite{man,zee},
\begin{align}\label{eq:g}
G_W(z,\bar{z})=\lim_{\lambda\to0,N\to\infty}\frac{1}{N}\mathrm{tr}_N\mathcal{G}_1,
\end{align}
where the limit $N\to\infty$ is taken before $\lambda\to0$. Similarly, as in Eqs.\eqref{eq:rho}-\eqref{eq:g}, 
we define $\mathcal{G}_{J\Lambda}$, $\Sigma_{J\Lambda}$ and $G_{J\Lambda}$. Introducing a constant matrix
\begin{align}\label{eq:m}
\mathcal{M}=
\left( \begin{array}{cc} M &  \\  & M^T  \end{array}\right),
\end{align} 
we have the relation
\begin{align}\label{eq:relat}
\mathcal{G}_{W}(\mathcal{Z})=\mathcal{G}_{J\Lambda}(\mathcal{Z}-\mathcal{M})
=\frac{1}{\mathcal{Z}-\mathcal{M}-\Sigma_{J\Lambda}}. 
\end{align}
It is impossible to calculate $\Sigma_{J\Lambda}$ explicitly for arbitrary $V$. But for our purpose it is 
sufficient to know its basic structure. Without loss of generality, assume $V(x)=x/2+\dots$, so that we 
can expand $\mathrm{tr}V(JJ^T)$ as 
\begin{align}\label{eq:exp}
\mathrm{tr}_NV(JJ^T)=\frac{1}{2}\mathrm{tr}_NJJ^T+g_2\mathrm{tr}_N(JJ^T)^2+g_3\mathrm{tr}_N(JJ^T)^3+\cdots.
\end{align} 
We expand the higher order terms in Eq.\eqref{eq:exp} and use the quadratic term to calculate the ensemble 
averages, denoted by $\langle\ \ \rangle_0$. Let $\bar{W}=J\Lambda$. Then because of the presence of matrix 
$\Lambda$, we have  
\begin{align}\label{eq:ru0}
\langle \bar{W}_{ab}\bar{W}^T_{cd}\rangle_0=\frac{\Lambda_{bb}^2}{N}\delta_{ad}\delta_{bc},\ \ \langle 
\bar{W}_{ab} \rangle_0=0.
\end{align}

\begin{figure}[!htb]
\centering
\begin{tikzpicture}[scale=3]
\draw (-1.53,0) -- (-0.94,0) (-0.66,0)-- (-0.07,0);
\draw (0,0) arc (0:78.5:8mm);  \draw (-0.07,0) arc (0:77:7.3mm);
\draw (-1.6,0) arc (180:100.5:8mm);\draw (-1.53,0) arc (180:103:7.3mm);
\draw (-0.8,0.72) circle (1.7mm); \draw (-0.8,0) circle (1.4mm);\draw (-0.79,0) node {$\mathcal{G}_{J\Lambda}$};
\draw (-0.8,0.72) node {$\Gamma_2$};
\draw (0.1,0) node {$+$};
\draw (1.8,0) arc (0:78.5:8mm);  \draw (1.8-0.07,0) arc (0:77:7.3mm);
\draw (1.8-1.6,0) arc (180:100.5:8mm);\draw (1.8-1.53,0) arc (180:103:7.3mm);
\draw (1.8-0.8,0.72) circle (1.7mm); 
\draw (0.71,0) arc (180:140:9mm);\draw (0.78,0) arc (180:138:8.3mm);
\draw (1.29,0) arc (0:40:9mm);\draw (1.22,0) arc (0:42:8.3mm);
\draw (0.49,0) circle (1.4mm);\draw (0.49,0) node {$\mathcal{G}_{J\Lambda}$};
\draw (1,0) circle (1.4mm);\draw (1,0) node {$\mathcal{G}_{J\Lambda}$};
\draw (1.51,0) circle (1.4mm);\draw (1.51,0) node {$\mathcal{G}_{J\Lambda}$};
\draw (0.27,0) -- (0.35,0) (0.63,0)-- (0.71,0) (0.78,0) -- (0.86,0) (1.14,0) -- (1.22,0) 
(1.29,0) -- (1.37,0) (1.65,0) -- (1.73,0);
\draw (1,0.72) node {$\Gamma_4$};
\draw (2,0) node {$+\cdots$};
\draw (-1.85,0 ) node {$\Sigma_{J\Lambda}=$};
\end{tikzpicture}
\caption{Contributions of the quadratic cumulant $\Gamma_2$ and quartic cumulant $\Gamma_4$ to the 
self-engergy $\Sigma_{J\Lambda}$.  }
\label{fig:sigma}
\end{figure}
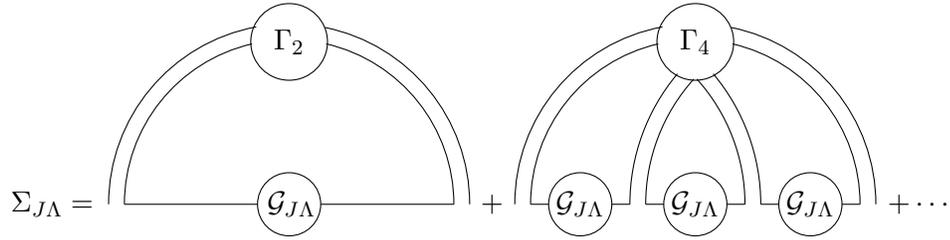

The self-energy $\Sigma_{J\Lambda}$ can be written in terms of the cumulants of $P(J)$, i.e. $\Gamma_{2k}$, $k=1,2,\dots$.
And it is well known that 
in the limit $N\to\infty$, to leading order in $\frac{1}{N}$, each of these cumulants is the sum of all connected 
plannar diagrams with $k$ external $J$'s and $J^T$'s. All diagrams which contribute to $\Sigma_{J\Lambda}$ are also
planar diagrams, as shown in Fig.\ref{fig:sigma}. By Eq.\eqref{eq:ru0}, the self-energy $\Sigma_{J\Lambda}$ has 
the following structure
\begin{align}\label{eq:sg}
\Sigma_{J\Lambda}=\left(\begin{array}{cc} \Sigma_1 & \Sigma_2 \\  \Sigma_3 & \Sigma_4\end{array}\right)
=\frac{1}{N}\left(\begin{array}{cc} 0 & a I_N \\ 
b \Lambda^2 & 0\end{array}\right),
\end{align} 
where scalars $a$ and $b$ are functions of $z$ and $\bar{z}$ and are determined by $V$ and $\Lambda$ is defined 
in Eq.\eqref{eq:la}. In appendix A, we prove that $G_{W}=G_{J\Lambda}$. This fact, together with Eq.\eqref{eq:rho}, 
completes the proof that the eigenvalue spectrum of $W$ is identical to that of the random matrix $J\Lambda$, 
as discovered in \cite{ra} when $J$ belongs to Ginibre ensemble. 

\begin{figure}[\!htb]
\centering
\includegraphics[scale=.55]{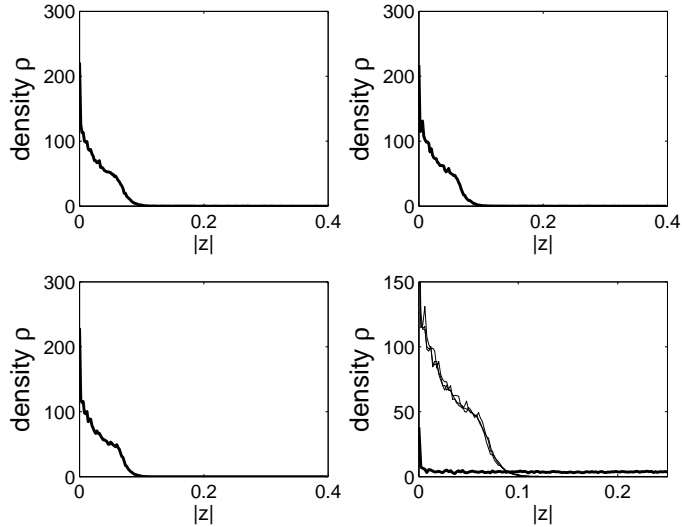}
\caption{Density $\rho$ of eigenvalues as a function of radius $|z|$ in the complex plane. Simulations are run 
for random matrices of dimension $N=200$, drawn from the ensemble defined in Eq.\eqref{eq:ensemble} with $V(x)=x+x^2$ . 
There are four types of neurons in the network, with $f=(0.1,0.2,0.3,0.4)$ and 
$\Lambda=\diag(0.5I_{20},1.0I_{40},1.5I_{60},2.0I_{80})$. Panel 1, $\mu=(10,30,30,-40)$. Panel 2, $\mu=(1,3,3,-4)$. 
Panel 3, $\mu=(0,0,0,0)$, i.e. $M=0$. Panel 4, density functions in panel 1-3 are drawn in the same plane. We 
find the bulk of the three 
functions are very similar. Which shows that even when elements of weight matrix $M$ are of order higher than 
$1/\sqrt{N}$, density function of $J\Lambda+M$ still converges to that of $J\Lambda$ as $N\to\infty$. This is due 
to the column structure of $M$. In comparison, we show in panel 4 the density function of $W=J\Lambda+M$, where $M$ 
is a constant matrix whose elements are randomly chosen from uniform distribution on $[0,1]$.
}
\label{fig:digraph}
\end{figure}

In Fig.\ref{fig:digraph}, we compare the eigenvalue spectra of $W=J\Lambda+M$ and 
$W=J\Lambda$. In both cases $J$ is drawn from a non-Gaussian ensemble. All spectra are generated by Monte-Carlo 
simulations. Since $\rho(x,y)=\rho(|z|)$, where $z=x+\mathrm{i}y$, it is sufficient to show the dependence of 
eigenvalue density function on radius $|z|$. We find these functions match quite well. In comparison, we 
replace $M$ with a constant matrix which does not have the column structure,
and find the density function is rather different.

Next, we test our result on the two-component Gaussian network of \cite{ra}. Denote the weight matrix by $W=J\Lambda+M$ 
and follow the brief discussion in the introduction. We fix $f\!=\!0.2$ and 
choose the variance of the Gaussian distributions for excitatory and inhibitory synaptic strengths to be 
$\frac{2}{N}$ and $\frac{1}{N}$, respectively. The only constraint on the mean strength matrix $M$ is the 'balance' 
condition. As in the previous example and already pointed out in \cite{ra}, the bulk of the spectra of $W$ with different 
$M$ matrices are almost identical. From numerical simulations, it appears that when $M$ has larger elements, there 
are more eigenvalues outside the support of the spectrum, see Eq.{\eqref{eq:bound}} for the definition. To show this 
is a finite-size effect, we calculate the percentage of 'outliers', for different matrix size $N$. It shows in 
Fig.{\ref{fig:lines}} that, for all values of $M$, as $N\to\infty$, percentages of 'outliers' for $W=J\Lambda+M$ approach 
to that of $W=J\Lambda$.

\begin{figure}[h]
\centering
\includegraphics[scale=.5]{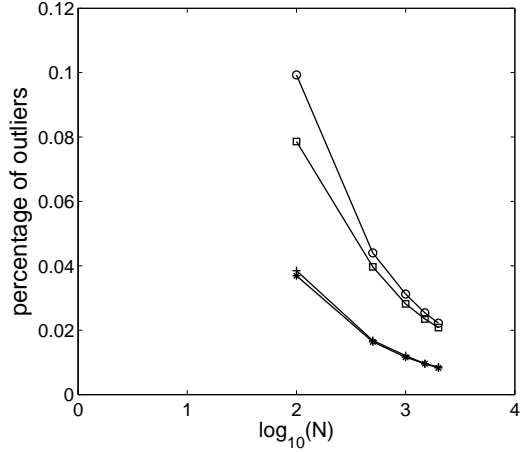}
\caption{Percentage of eigenvalue outliers of two-component Gaussian network $W$ decays as $N$ increases. 
For $f\!=\!0.2$, elements in $fN$ columns of $W$ have Gaussian distribution with variance $\frac{2}{N}$ and 
mean $\mu_e$. Other elements have Gaussian distribution with variance $\frac{1}{N}$ and mean $\mu_i$. From top 
to bottom,  $(\mu_e,\mu_i)$ are $(1,-\frac{1}{4})$, $(\frac{1}{\sqrt{N}},-\frac{1}{4\sqrt{N}})$, 
$(\frac{1}{N},-\frac{1}{4N})$ amd $(0,0)$, respectively. }
\label{fig:lines}
\end{figure}


\section{Synaptic strength of Gaussian distribution}

In this section, we calculate the eigenvalue density of multi-component Gaussian network. Assume there are $m$ 
types of neurons in the network and synaptic strengths have different Gaussian distributions. From the previous 
section we know the density functions of $J\Lambda+M$ and $J\Lambda$ are identical when $M$ has the 
column structure and satisfies the 'balance' condition, even without the additional constraint of \cite{ra}.  
Therefore, spectrum of this network is the density function $\rho(x,y)$ of the following random matrix
\begin{align}
W=J\Lambda,
\end{align}
where $J$ is drawn from the real Ginibre ensemble, i.e. $V(x)=\frac{1}{2}x$ in Eq.\eqref{eq:ensemble}, 
and $\Lambda$ is defined in Eq.\eqref{eq:la}. The case $m=2$ 
is solved in \cite{ra} with the method in \cite{som}. For $m>2$, we find the technique developed in \cite{man,zee} 
is more convenient. Define an operator $\bar{\mathrm{tr}}_N$ which, when acts on an $N\times N$ matrix $A$, 
gives $\bar{\mathrm{tr}}_NA=\mathrm{tr}_NA\Lambda^2$. By Eq.\eqref{eq:ru0}, the equation for the one particle 
irreducible (1PI) self-energy $\Sigma_W$ is
\begin{align}\label{eq:sd}
\Sigma_W=\left(\begin{array}{cc} \Sigma_1 & \Sigma_2 \\  \Sigma_3 & \Sigma_4\end{array}\right)
=\frac{1}{N}\left(\begin{array}{cc} 0 & \bar{\mathrm{tr}}_N\mathcal{G}_2\cdot I_N \\ 
{\mathrm{tr}}_N\mathcal{G}_3\cdot \Lambda^2 & 0\end{array}\right).
\end{align} 
Note that the above matrix has the structure outlined in Eq.\eqref{eq:sg}. 
The generalized Green's function $\mathcal{G}_W$ (defined in Eq.\eqref{eq:G}) is related to $\Sigma_W$ by 
the Schwinger-Dyson equation Eq.\eqref{eq:self},
\begin{align}\label{eq:gg}
\mathcal{G}_W=
\left(\begin{array}{cc} \mathcal{G}_1 & \mathcal{G}_2 \\ \mathcal{G}_3 & \mathcal{G}_4\end{array}\right)=
\left( \begin{array}{cc} z-\Sigma_1 & \lambda-\Sigma_2 \\ 
\lambda-\Sigma_3 & \bar{z}-\Sigma_4 \end{array} \right)^{-1}_{2N\times2N}.
\end{align}

\begin{figure}[!htb]
\centering
\begin{tikzpicture}[scale=3]
\draw (0,0) arc (0:180:10mm); 
\draw (-1.93,0) arc (180:0:9.3mm); 
\draw (-1,0) circle (2mm);
\draw (-1.93,0) -- (-1.2,0); 
\draw (-0.8,0)-- (-0.08,0);
\draw (-1,0) node {$\mathcal{G}_W$};
\draw (-2.25,0 ) node {$\Sigma_W=$};
\end{tikzpicture}
\caption{Self-energy $\Sigma_W$ is related to resolvent $\mathcal{G}_W$ by equation Eq.\eqref{eq:sd}.}
\label{fig:past}
\end{figure}
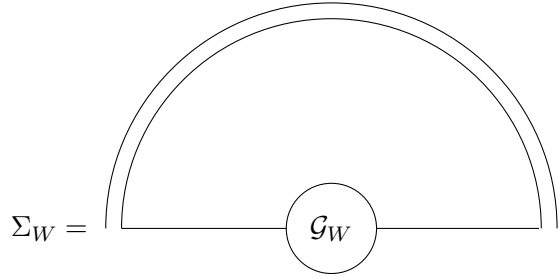
 
From Eqs.\eqref{eq:sd} and \eqref{eq:gg}, we get the following equation for an unknown variable 
$p=\frac{1}{N}\bar{\mathrm{tr}}_N\mathcal{G}_2\cdot \frac{1}{N}\mathrm{tr}_N\mathcal{G}_3$,
\begin{align}\label{eq:y}
\sum_{i=1}^m\frac{f_i\sigma_i^2}{|z|^2-p\sigma_i^2}=1.
\end{align}
Eq.\eqref{eq:y} has multi-number of solutions. The correct one for our problem is the one satisfying the boundary 
condition
\begin{align}\label{eq:bd}
p|_{|z|^2=0}=-1.
\end{align}
The boundary of spectrum is determined by the transition point $p=0$ \cite{man}, which corresponds to the circle 
with radius $|z|_B$, such that
\begin{align}\label{eq:bound}
|z|^2_B=\sum_{i=1}^{m}f_i\sigma_i^2.
\end{align}
The disk region defined by $|z|\le|z|_B$ is the support of the spectrum. Off the support, we always have $p=0$. 
In the case $m=2$, the formula in Eq.\eqref{eq:bound} gives the same result for spectrum boundary obtained 
in \cite{ra} by solving 
a saddle-point equation. From \eqref{eq:g}, the Greens's function $G_W(z,\bar{z})$ 
is given by the following formula 
\begin{align}\label{eq:greenjs}
G_W(z,\bar{z})=\left\{
\begin{array}{ll} \bar{z}\sum_{i=1}^m\frac{f_i}{|z|^2-p\sigma_i^2},& |z|^2\le|z|^2_B \\ & \\
\frac{1}{z}, & |z|^2>|z|^2_B
\end{array}\right. .
\end{align}
From Eq.\eqref{eq:y}, when on the support of spectrum, we get 
\begin{align}\label{eq:dydz}
\frac{\partial p}{\partial |z|^2}=\frac{\sum_{i=1}^m\frac{f_i\sigma_i^2}{(|z|^2-p\sigma_i^2)^2}}
{\sum_{i=1}^m\frac{f_i\sigma_i^4}{(|z|^2-p\sigma_i^2)^2}}.
\end{align}
Finally, by Eq.\eqref{eq:rho}, we get the eigenvalue density of $W$
\begin{align}\label{eq:so}
\rho(x,y)=\left\{\begin{array}{ll}
\frac{1}{\pi}\left(|z|^2\frac{\partial p}{\partial |z|^2}-p\right)
\sum_{i=1}^m\frac{f_i\sigma_i^2}{(|z|^2-p\sigma_i^2)^2},& |z|^2\le|z|^2_B \\ & \\
0, & |z|^2>|z|^2_B\end{array}\right. .
\end{align}
Introduce the notation $\langle\sigma^a\rangle=\sum_i^m f_i\sigma_i^a$, where $a$ is a constant. 
From Eq.\eqref{eq:so}, we find the eigenvalue density at the centre and boundary of the spectrum
\begin{align}
\rho(0)=\frac{1}{\pi}\langle\sigma^{-2}\rangle, \ \ {\mathrm{and}}\ \ 
\rho(|z|_B)=\frac{1}{\pi}\frac{\langle\sigma^2\rangle}{\langle\sigma^4\rangle}.
\end{align}

\begin{figure}[!htb]
\centering
\includegraphics[scale=.42]{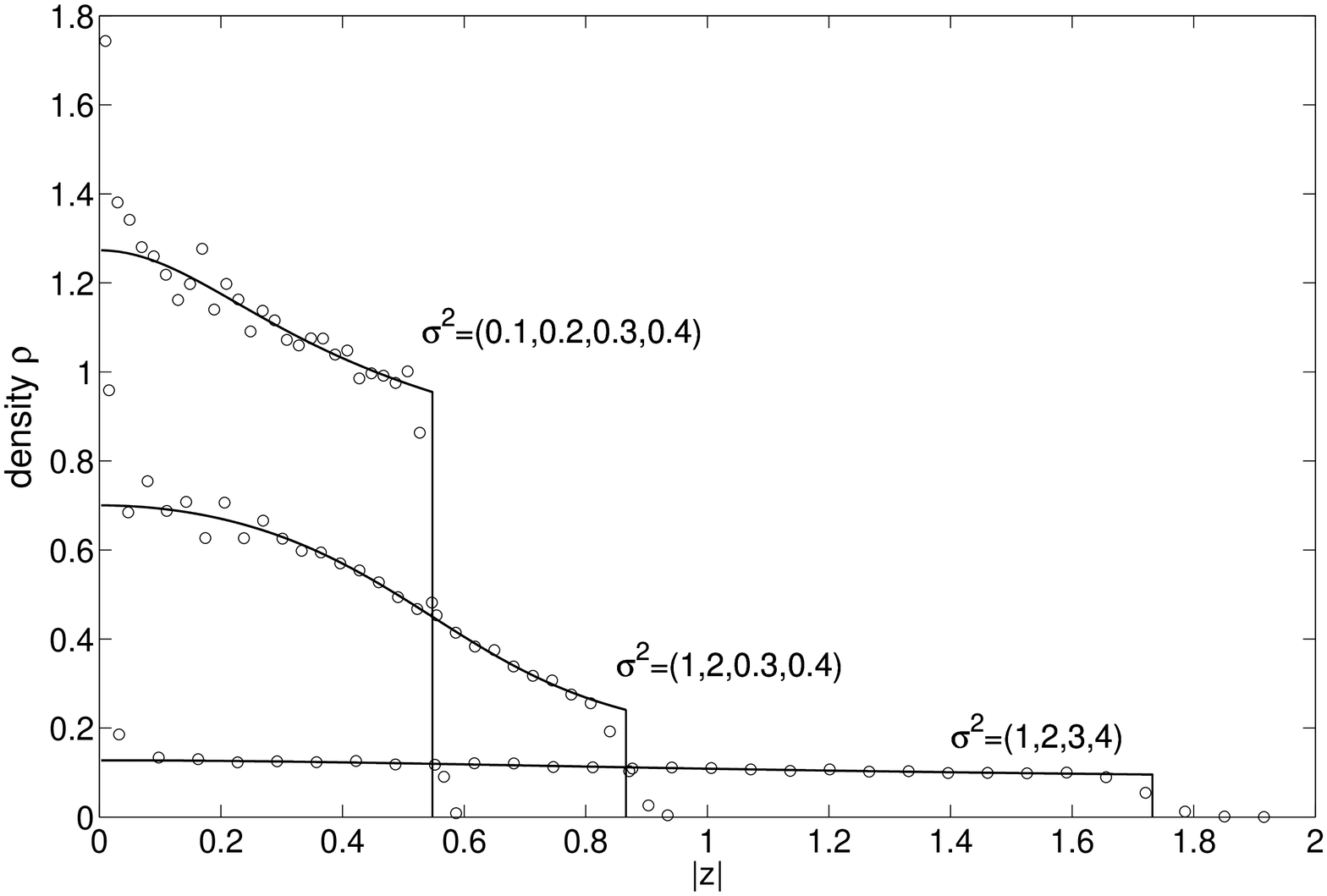}
\caption{Density $\rho$ of eigenvalues as a function of radius in the complex plane $|z|$, for $N=400$. 
The solid lines are the analytic results by Eq.\eqref{eq:so} and symbols are numerical 
simulations. The figure shows results for different sets of variances $\sigma^2/N$ with fixed population 
$f=(0.1,0.2,0.3,0.4)$.}
\label{fig:digraph1}
\end{figure}
 
\begin{figure}[!htb]
\centering
\includegraphics[scale=.42]{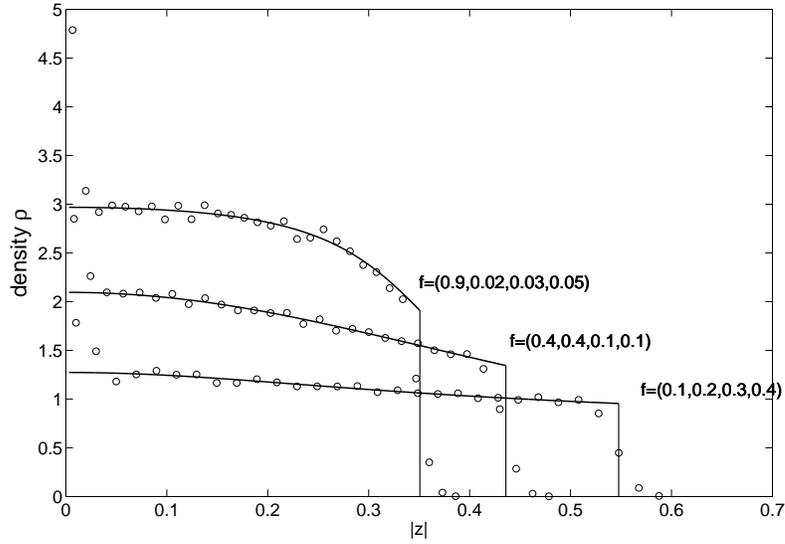}
\caption{Density $\rho$ of eigenvalues as a function of radius in the complex plane $|z|$, for $N=400$. 
The solid lines are the analytic results by Eq.\eqref{eq:so} and symbols are numerical 
simulations. The figure shows results for different sets of population with fixed variances $\sigma^2=(0.1,0.2,0.3,0.4)/N$.}
\label{fig:digraph2}
\end{figure}

When $m=1$, from Eqs.\eqref{eq:y} and \eqref{eq:so} we easily recover the well known result for Ginibre ensemble \cite{gin}. 
For $m=2$, choosing the solution for quadratic equation Eq.\eqref{eq:y} satisfying condition Eq.\eqref{eq:bd}, then from 
Eq.\eqref{eq:so}, we 
successfully recover the results obtained in \cite{ra}. For large $m$, it is hardly possible to have an analytic 
solution for Eq.\eqref{eq:y}. But it is 
very simple to find the numerical solution for this algebraic equation. It turns out that there is always only 
one solution on $[-1,0]$, which is just what we need according to the boundary conditions.   

In Fig.\ref{fig:digraph1}, we compare the density function in Eq.\eqref{eq:so} with numerical simulations 
for synaptic strengths with different variances $\sigma$'s but the same $f$'s. In Fig.\ref{fig:digraph2}, we let $f$'s change 
but keep $\sigma$'s fixed. In both cases, we observe very good match between numeric data and analytical results. The 
only significant deviation happens near $|z|=0$. In fact, this deviation already appears when $\Lambda=I_N$ and is shown 
due to finite-size effect \cite{som}.

\section{Discussion}

In the first part of this paper, we show that modifying the mean strengths of synapses of a neural network does 
not change the density function of synaptic matrix
even when there are several types of neurons and the 
strengths of their synaptic connections have correlated distributions.

In Eq.\eqref{eq:ensemble}, the ensemble of random matrix $J$ is chosen to be $O(N)$ invariant so that 
all elements of random matrix $J$ have the same distribution. Differences between different types of neurons are 
introduced only by $\Lambda$ and $M$. In fact, we can draw the synaptic matrix $W$ from more general ensembles. 
As long as Eq.\eqref{eq:ru0} holds and $M$ has the column block structure, eigenvalue 
spectra of $W$ will not depend on $M$.  

We therefore prove that the density functions of large random matrices described by 
Eq.\eqref{eq:w}-\eqref{eq:ensemble} are not changed by perturbations of the rank-1 matrix $M$. 
This type of random matrices are not of Wigner type or have iid elements as in \cite{tt,ss}.

It is its structure that makes $M$ irrelevant to the eigenvalue density function. In reality, we may need to choose 
the mean value of synaptic connections to be of the same order of their fluctuations, 
i.e. $\langle M_{ij}\rangle\propto N^{-1/2}$. 
But this is not necessary in our proof. If the 'balance' condition is not imposed, the eigenvalue spectra will be 
identical to the 'balanced' case except the eigenvalues at zero will be shifted \cite{ra,tt}. 
  
In the second part of this paper we calculate the density function of random matrices of the form $J\Lambda$, where $J$ 
belongs to Ginibre ensemble. These matrices describe random networks with multiple independent components. We find 
closed formulas for the eigenvalue density 
at both the centre and the boundary of the spectrum in terms of variances of synaptic strengths. 

When $J$ is drawn from the ensemble in Eq.\eqref{eq:ensemble} and $\Lambda=I_N$, we know by the Single-Ring Theorem 
\cite{fz,frz,f} that the support of the eigenvalue spectrum is either a disk or an annulus. It will be interesting 
to find out whether or not the Single-Ring Theorem still holds when $\Lambda$ is diagonal but not proportional to 
$I_N$. Eq.\eqref{eq:bound} shows when $J$ has Gaussian distribution the support of the spectrum is always a disk of 
radius $|z|_B$, but never an annulus. This indeed agrees with the Single-Ring Theorem. Clearly, to prove the 
Single-Ring Theorem for $J\Lambda$, where $J$ belongs to the general ensemble defined in Eq.\eqref{eq:ensemble},  
we need to take different approaches. Work on this topic is currently in process. 


\section*{Acknowledgement}
I am grateful to Prof. R.T. Scalettar for helpful communication on numerical methods and giving me his Monte-Carlo 
code. I also thank Dr. K. Rajan for drawing my interest to this research. This work was supported by the Swartz 
Foundation.

\section*{Appendix}
\appendix
\section{}
In this section we show that in the large $N$ limit, due to the structure of $\mathcal{M}$, the Green's 
function $G_W$ defined in Eq.\eqref{eq:g} equals to $G_{J\Lambda}$ defined similarly for $J\Lambda$, i.e. 
\begin{align}\label{eq:ap}
G_W=G_{J\Lambda}+O(N^{-1}).
\end{align}

Step 1. Define $D=(\bar{z}-M^T)(z-M)-ab\Lambda^2$, where $a$ and $b$ are scalars shown in Eq.\eqref{eq:sg}. 
By Eq.\eqref{eq:relat},  $\mathcal{G}_W$ can be written as 
\begin{align}\label{eq:ginv}
\mathcal{G}_W=
\left(\begin{array}{cc}D^{-1}(\bar{z}-M^T) & a D^{-1} \\ * & (z-M)D^{-1}\end{array} \right).
\end{align}
Here we used the fact that the (12)-element of $\Sigma_{J\Lambda}$ is proportional to identity matrix and the following 
formula from linear algebra
\begin{align}\label{eq:inv}
\left(\begin{array}{cc} E & F \\ G & H\end{array} \right)^{-1}=
\left(\begin{array}{cc}E^{-1}+E^{-1}FX^{-1}GE^{-1} & -E^{-1}FX^{-1} \\ -X^{-1}GE^{-1} & X^{-1}\end{array} \right),
\end{align}
where $X=H-GE^{-1}F$.

Step 2. Let $\mathbb{I}_{pq}$ be a $p\times q$-dimensional matrix with all elements equal to 1 and 
let $M_i=f_iN$ for $i=1,\dots,m$. Then $D^{-1}$ has the following $m\times m$-block structure
\begin{align}\label{eq:dinv}
D^{-1}=\left(\begin{array}{cccc}
a_1+b_{11}\mathbb{I}_{M_1M_1} & b_{12}\mathbb{I}_{M_1M_2} & \cdots & b_{1m}\mathbb{I}_{M_1M_m}  \\ 
b_{21}\mathbb{I}_{M_2M_1} & a_2+b_{22}\mathbb{I}_{M_2M_2} & \cdots  & b_{2m}\mathbb{I}_{M_2M_m} \\
\vdots & \vdots & \ddots & \vdots \\
b_{m1}\mathbb{I}_{M_mM_1} & b_{m2}\mathbb{I}_{M_mM_2} & \cdots &  a_m+b_{mm}\mathbb{I}_{M_mM_m}
\end{array} \right),
\end{align} 
where $a_i=1/(|z|^2-ab\lambda^2_i)$, $b_{ii}=O(N^{-1})$, $b_{i\ne j}=O(N^{-2})$, and $a_i+b_{ii}M_i=O(N^{-1})$, 
for $i,j=1,\dots,m$. This claim is proved by induction. First, notice that $D$ has the same $m\times m$-block 
structure as in \eqref{eq:dinv} except all its parameters are of order 1. Using the fact 
\begin{align}\label{eq:id}
(a+b\mathbb{I}_{MM})^{-1}=\frac{1}{a}-\frac{b}{a(bM+a)}\mathbb{I}_{MM},
\end{align}
and \eqref{eq:inv}, we find that for $m=2$, $D^{-1}$ indeed has the properties described by \eqref{eq:dinv}. 
Then assume the claim is true for $m=n-1$. By straightforward calculation using \eqref{eq:inv} and 
\eqref{eq:id}, we find the claim is also true for $m=n$. 

Step 3. Substituting \eqref{eq:dinv} to \eqref{eq:ginv} and using Eq.\eqref{eq:g}, we get \eqref{eq:ap}. 
This completes the proof.

\end{document}